# Unlocking ultrafast spin dynamics in a rare-earth magnet

Paul Herrgen,[1,*] Christian Holzmann,[2] Torben Manzke,[1] Ulrich Nowak,[3] Peter M. Oppeneer,[4] Manfred Albrecht,[2] Benjamin Stadtmüller,[1,2] and Martin Aeschlimann[1]

[1]*Department of Physics and Research Center OPTIMAS, RPTU University Kaiserslautern-Landau, 67663 Kaiserslautern, Germany*
[2]*Institute of Physics, University of Augsburg, 86159 Augsburg, Germany*
[3]*Department of Physics, University of Konstanz, 78457 Konstanz, Germany*
[4]*Department of Physics and Astronomy, Uppsala University, 75120 Uppsala, Sweden*

The speed of optically driven magnetization dynamics is fundamentally determined by how efficiently angular momentum can be transferred between electronic, spin and lattice degrees of freedom. In rare-earth magnets, this process is typically slow because optical excitation primarily addresses itinerant electrons, whereas the magnetic moment resides in localized 4f states. Here we show that selective optical excitation of localized magnetic states can overcome this limitation. Using femtosecond pump–probe magneto-optical spectroscopy of ferrimagnetic gadolinium iron garnet, we resonantly excite an intra-4f transition of $Gd^{3+}$ at 4.65 eV and resolve the ensuing dynamics of the antiferromagnetically coupled Gd and Fe sublattices. Direct excitation of the 4f manifold induces an ultrafast demagnetization of the Gd sublattice with a characteristic time of 38 fs, more than two orders of magnitude faster than in elemental gadolinium and even faster than the response of the Fe sublattice in the same material. By contrast, off-resonant excitation strongly suppresses the acceleration of the Gd dynamics while leaving the Fe response largely unchanged. These results demonstrate that the ultrafast magnetic response of rare-earth systems is governed not only by intrinsic material properties but also by the optical excitation pathway. Selective access to localized magnetic states therefore provides a powerful photonic handle for engineering angular-momentum flow and controlling magnetism far from equilibrium.

## INTRODUCTION

The ability to manipulate magnetic order on ultrafast timescales has fundamentally reshaped our understanding of spin dynamics in solids. Since the first observations of femtosecond laser-induced demagnetization [1], it has become evident that magnetic systems can exchange angular momentum with their environment far more rapidly than anticipated from equilibrium considerations. Yet, despite decades of research, the microscopic pathways that govern these dynamics - and their strong material dependence - remain a central open question.

A particularly revealing contrast is found between transition-metal and rare-earth magnets. In itinerant 3d ferromagnets such as Fe, Co and Ni, demagnetization occurs on $100 \pm 50$ fs timescales, enabled by efficient coupling between spin, charge and lattice degrees of freedom via spin-orbit interaction [2–5]. By contrast, rare-earth systems like gadolinium (Gd) and terbium (Tb) exhibit significantly slower demagnetization dynamics, often extending into the tens of picoseconds [6, 7]. This striking difference reflects the dual nature of magnetism in rare-earth elements: while optical excitation predominantly addresses itinerant 5d and 6s electrons, the magnetic moment resides in localized 4f states. As a result, the laser-induced energy transfer from the conduction electrons (5d and 6s) to the localized 4f magnetic moments is inefficient and proceeds via slow, indirect pathways, giving rise to the observed long demagnetization times [8]. The resulting indirect energy transfer between electronic subsystems imposes a severe bottleneck on the spin dynamics of rare-earth elements.

The sensitivity of ultrafast demagnetization to electronic structure is further highlighted by comparisons within the rare-earth series. Gd and Tb exhibit a two-step demagnetization process [4], featuring an ultrafast component with a time constant of 750 fs, identical in both cases [7]. However, the slower demagnetization times differ significantly: 40 ps for Gd and 8 ps for Tb. In Gd, which has a half-filled 4f shell and vanishing orbital angular momentum ($L = 0$), the spin–orbit interaction is very weak, resulting in an effective decoupling of spin and orbital degrees of freedom. In contrast, Tb possesses a finite orbital moment ($L = 3$), leading to stronger spin–orbit coupling and enabling a demagnetization process approximately five times faster than in Gd. These observations indicate that the efficiency of angular momentum transfer - and consequently the speed of magnetization dynamics - is governed by both the specific excitation pathways that couple the optical field to the magnetic degrees of freedom and the strength of the spin–orbit interaction.

An alternative route to accelerating spin dynamics in rare-earth systems is provided by ferrimagnetic compounds, where rare-earth and transition-metal sublattices are antiferromagnetically coupled. In such systems, additional spin-flip channels for angular momentum exchange emerge through inter-sublattice spin-spin exchange scattering [9–13].

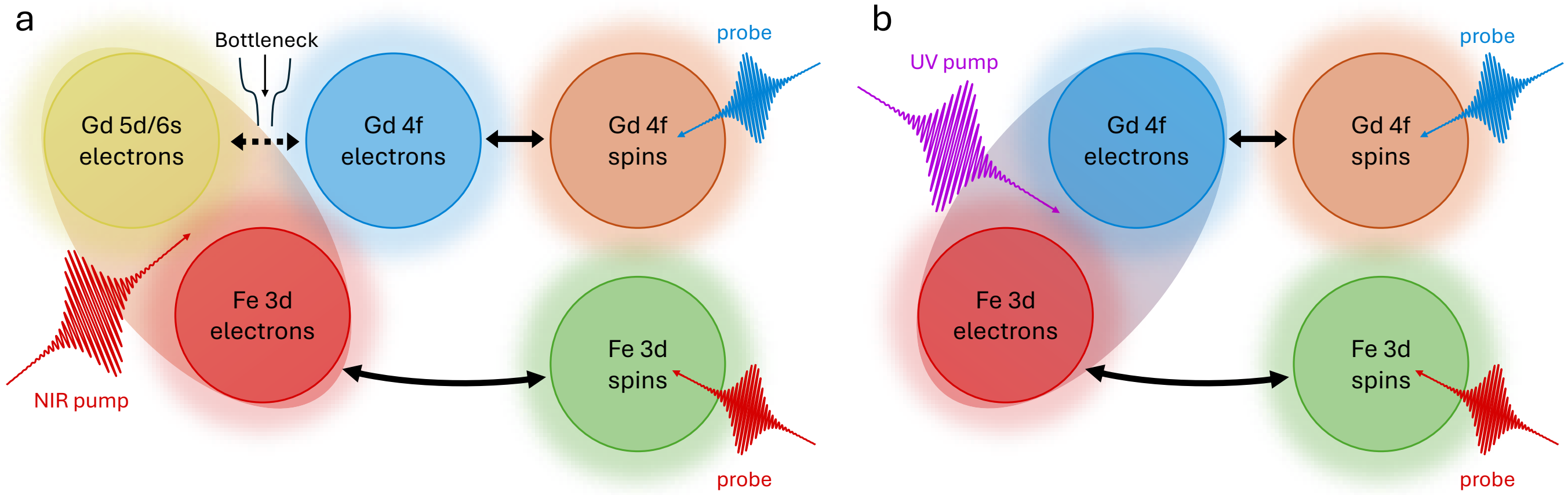

FIG. 1: (**a**) Conventional NIR excitation couples to itinerant 5d/6s (Gd) and 3d (Fe) electrons. The slow coupling to the Gd 4f electrons creates a bottleneck, leading to slow magnetization dynamics. (**b**) Resonant 4f-4f excitation using UV pulses addresses the magnetic Gd 4f electrons directly and bypasses the bottleneck, resulting in a much faster demagnetisation process.

In Gd–Fe alloys, this mechanism leads to rich non-equilibrium phenomena, including ultrafast switching and transient ferromagnetic states [14]. Nevertheless, a pronounced asymmetry persists: while the 3d transition-metal sublattice responds on 100 fs timescales, the Gd 4f moments remain still comparatively slow (in the ps timescale), reflecting their indirect coupling to the optical excitation.

This raises a fundamental question: to what extent are the ultrafast dynamics of localized magnetic moments limited by intrinsic material properties, and to what extent by the nature of the excitation?

In itinerant 3d ferromagnets, it was shown that magnetization dynamics remain nearly identical across a range of pump photon energies [5] and polarizations [15]. By contrast, due to the dual nature of magnetism in rare-earth elements (see above), Gd and Tb should exhibit demagnetisation dynamics that critically depend on the excitation mechanism.

However, in nearly all published experiments, the optical pump - typically either the fundamental (1.55 eV) or second harmonic (3.1 eV) of a Ti:sapphire laser - couples predominantly to delocalized electronic states [8, 14, 16–20]. As a result, the localized 4f moments are only indirectly perturbed, see Fig. 1a. The ensuing dynamics are therefore governed by energy and angular momentum transfer between weakly coupled subsystems as discussed above. A direct optical excitation of the 4f electrons, in principle, would circumvent this bottleneck and could provide a qualitatively different pathway for ultrafast spin control.

Here we demonstrate that such a regime can be realized experimentally. We investigate a ferrimagnetic garnet, GdIG, in which the Gd and Fe sublattices are antiferromagnetically coupled. By tuning the excitation to the intra-4f transition of Gd at approximately (4.65 eV), well above the (2.86 eV) bandgap [21], we directly address the localized magnetic 4f states. This excitation pathway fundamentally alters the non-equilibrium dynamics: instead of relying on indirect coupling via itinerant electrons, the magnetic subsystem is driven at its origin, see Fig. 1b.

We find that this approach leads to a drastic acceleration of the Gd spin dynamics. The demagnetization time of the Gd sublattice is reduced by orders of magnitude compared to elemental Gd and, remarkably, becomes even shorter than that of the Fe sublattice in the same material. This inversion of the conventional hierarchy of timescales demonstrates that the slow response of Gd is not an intrinsic limitation of its 4f magnetism, but rather a consequence of the excitation pathway.

Our results establish that ultrafast spin dynamics can be controlled not only through material design, but also through selective excitation of specific electronic subsystems. In this view, the flow of angular momentum is governed by the interplay between electronic structure, coupling mechanisms and optical selection rules. Direct access to localized magnetic states provides a powerful means to manipulate this flow, and enables spin dynamics regimes that cannot be achieved with the excitation wavelengths used to date.

## EXCITATION: EXPERIMENTAL EVIDENCE FOR AN INTRA-ATOMIC 4F-4F TRANSITION

In our time-resolved demagnetization experiment, we directly excite the 4f electrons via resonant laser excitation at the lowest $Gd^{3+}$ 4f–4f transition (270 nm) in a GdIG (gadolinium iron garnet) thin film. This well-established ∼4.65 eV transition, first characterized in early rare-earth spectroscopy studies [22], corresponds to a multiplet (intra-4f) excitation within the $4f^7$ manifold. In contrast, the exchange splitting between majority and minority 4f spin states is approximately ∼11 eV [23].

This resonant excitation is inherently a many-body, atomic-like process:

$$^8S_{\frac{7}{2}} \rightarrow\ ^6I_J \text{ with } J = \frac{7}{2}, \frac{9}{2}, \frac{11}{2}, \dots \frac{17}{2} \tag{1}$$

rather than a single-particle interband transition. Here, $J$ is the total angular momentum quantum number. Although these 4f–4f transitions are nominally spin-forbidden by both spin and parity selection rules, they acquire finite oscillator strength through a combination of weak symmetry breaking, crystal-field induced mixing of opposite-parity states (primarily 5d and ligand orbitals), and dynamic vibronic coupling. Experimentally, they manifest as a pronounced absorption feature, as recently confirmed by Pankratova *et al.* [24].

Optical excitations within the 4f shell of trivalent gadolinium ($Gd^{3+}$) embedded in crystalline hosts represent a particularly instructive class of localized electronic processes in rare-earth systems. In $Gd^{3+}$, the ground state configuration $4f^7$ corresponds to a half-filled shell with a high-spin $S = \frac{7}{2}$ state, which leads to an especially stable Hund's rule ground state. Excitations in this manifold must therefore involve transitions between strongly exchange-coupled, spin-aligned multiplet states. This leads to a rich fine structure and a strong dependence on host lattice symmetry and covalency.

## PROBING: WAVELENGTH TUNING TO PROBE INDIVIDUAL SUBLATTICE MAGNETIZATION DYNAMICS

To probe the influence of sublattice interactions on ultrafast demagnetization dynamics, it was essential to resolve the magnetization response of the Fe and Gd sublattices independently. This requires a method capable of distinguishing the contributions of each magnetic species, particularly in a ferrimagnetic system where the sublattices are coupled antiferromagnetically. To achieve this, we drew upon the theoretical framework proposed by Khorsand *et al.* [25] and confirmed by S. Alebrand *et al.* [26], who demonstrated that elemental-specific magneto-optical sensitivity can be attained in 3d–4f alloy systems using visible-light probing with distinct wavelengths. Their analysis indicates that the different magneto-optical response arises from the distinct electronic structures of the Fe 3d and Gd 4f electrons: the Fe sublattice, with its itinerant 3d electrons, exhibits a stronger response at longer wavelengths, while the Gd sublattice, dominated by localized 4f transitions, contributes more significantly at shorter wavelengths due to its resonant optical features. Based on this, we anticipated that a probe wavelength of 800 nm would exhibit enhanced sensitivity to the Fe sublattice, whereas a 400 nm probe would be more sensitive to the Gd sublattice. To test this prediction, we performed static magneto-optical hysteresis measurements using both probe wavelengths independently (see Fig. 2a). As expected, we observed a reversal in the sign of the hysteresis loop when switching from 800 nm to 400 nm, consistent with the antiparallel alignment of the Fe and Gd sublattices. This sign inversion provides direct evidence of the differential spectral sensitivity and confirms that the probe beam selectively monitors each sublattice. To further validate this interpretation, we conducted additional hysteresis measurements above and below the compensation temperature of GdIG ($T_{comp} \approx 240$ K), where the net magnetization vanishes and the relative magnetization of the two sublattices reverses due to the temperature-dependent dominance of one over the other. As shown in Fig. 2b, we observed a corresponding reversal in the hysteresis loop sign, confirming that the probe wavelength-dependent signal originates from distinct sublattice contributions. These results establish a robust experimental basis for disentangling the dynamics of the Fe and Gd sublattices and provide a reliable framework for interpreting ultrafast magneto-optical measurements in rare-earth-based ferrimagnets. This approach not only enables sublattice-resolved probing but also exemplifies the power of combining theoretical insight with targeted experimental validation to advance understanding in complex magnetic systems.

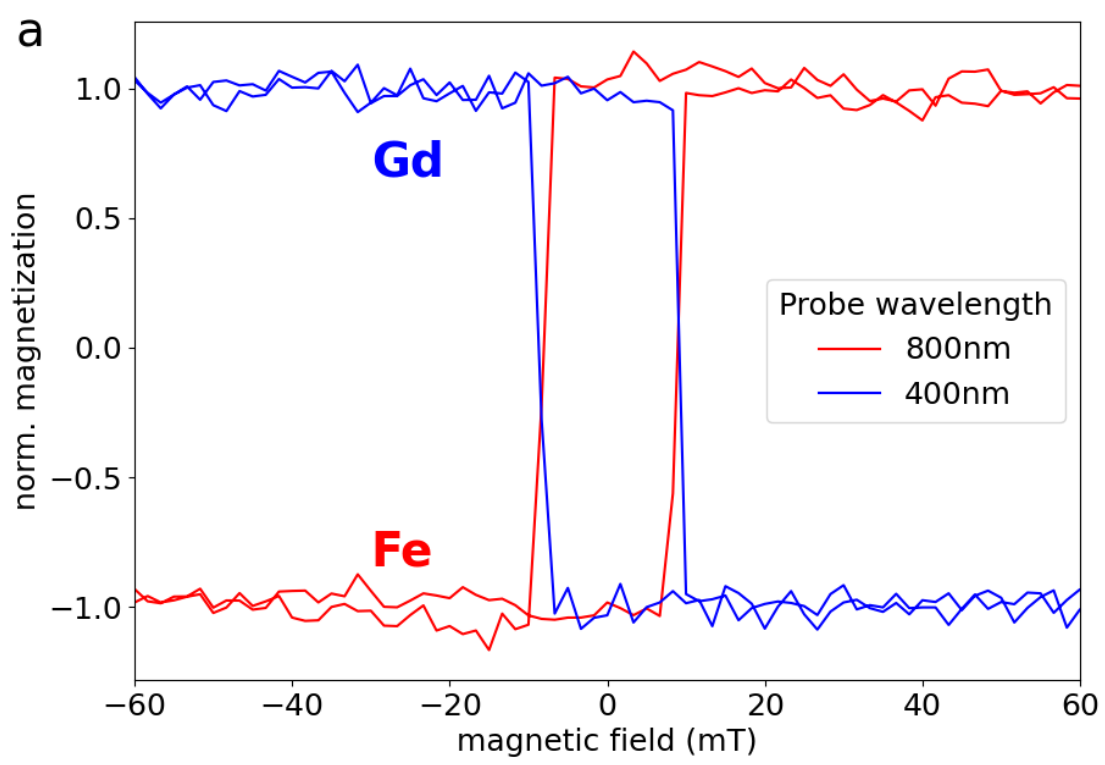

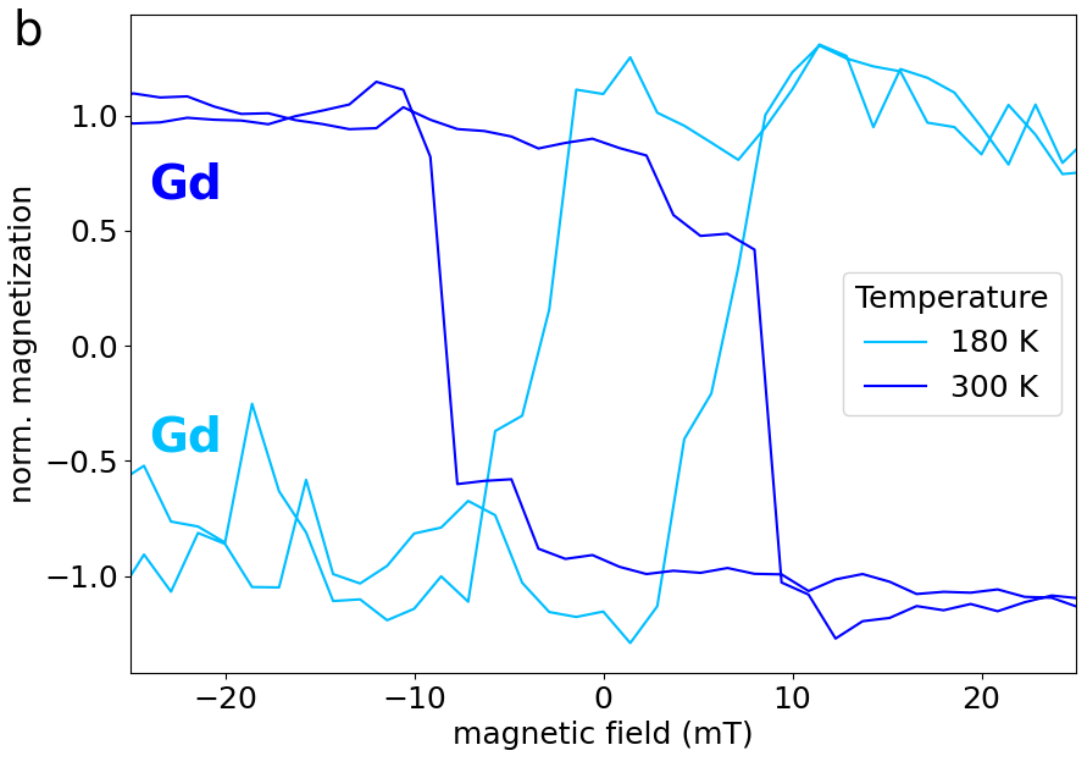

FIG. 2: Hysteresis measurements of both GdIG sublattices (**a**). The iron sublattice was measured with a 800 nm probe wavelength (red), the gadolinium sublattice was measured with a 400 nm probe wavelength (blue). (**b**) Hysteresis measurements of the Gd sublattice, measured below (light blue) and above (dark blue) the compensation temperature of GdIG. Both curves were measured using a 405 nm probe wavelegnth.

## RESULTS AND DISCUSSION

We performed time-resolved complex MOKE measurements to analyze the Kerr rotation and Kerr ellipticity of the Gd and Fe sublattices (see Methods). All measurements were carried out at room temperature, performed with a pump wavelength of 267 nm (4.65 eV), a pulse duration of 87 fs, and a pump power of $0.64\,\frac{\mathrm{mJ}}{\mathrm{cm}^2}$. The transient magnetization dynamics of both sublattices following resonant 4f–4f excitation at 267 nm are illustrated in Fig. 3 a. Immediately after optical excitation, both the Fe and Gd sublattices undergo ultrafast demagnetization, with remarkably similar initial dynamics within the first few hundred femtoseconds. Despite the final quenching levels differing significantly - reflecting the antiferromagnetic coupling and differing spin densities - their early-time evolution exhibits a striking temporal similarity, suggesting a shared mechanism driving the initial loss of magnetization. A double-exponential fit to the data yields demagnetization times $\tau_M$ of 59 fs for the Fe sublattice and 38 fs for Gd, both falling within the timescale characteristic of ultrafast demagnetization in 3d transition metal ferromagnets. The Fe demagnetization time of $\tau_M(\mathrm{Fe}) = 59$ fs is approximately three times faster than that observed in pure Fe films [27], but is in good agreement with values reported for ferrimagnetic GdFeCo alloy [14], indicating a strong influence of the antiferromagnetic coupling and interfacial spin dynamics. In contrast, the Gd sublattice exhibits a demagnetization time of $\tau_M(\mathrm{Gd}) = 38$ fs, which is significantly faster than the few tens of ps timescales reported for pure Gd films [16] or sub ps for ferrimagnetic GdFeCo alloy [14]. This discrepancy cannot be attributed to conventional spin-flip scattering or spin–orbit relaxation mechanisms alone, which are expected to operate similarly across both sublattices. Instead, the accelerated demagnetization of Gd suggests the presence of an additional, sublattice-specific mechanism that strongly couples to the 4f electrons.

Given the resonant excitation of the $4f^7$ multiplet states, we attribute the enhanced dynamics to an ultrafast change to excited 4f configurations with finite orbital angular momentum ($L \neq 0$). The result highlights the distinctive role of localized 4f excitations in driving ultrafast spin dynamics in rare-earth ferrimagnets and underscores the importance of sublattice-specific relaxation pathways.

The contrasting response of the two sublattices and the exceptionally fast Gd demagnetization provide clear evidence for a mechanism that goes beyond known mechanisms, such as magnon generation, electron-phonon angular momentum transfer, or a generic spin–orbit scattering decay channel [4, 27–31]. Instead, it depends sensitively on the electronic structure and optical excitation of the Gd 4f shell. We emphasize that these are not single-particle excitations but rather they involve multiple electrons within the $^6I_J$ multiplet of the $Gd^{3+}$ ion in GdIG. Upon excitation, the admixture of spin and orbital character renders spin no longer a good quantum number. In this regime, the weak spin-orbit interaction of the $L = 0$ ground state becomes effectively modified, promoting transfer from spin to orbital degrees of freedom. Antiferromagnetic exchange coupling to the Fe sublattice provides an additional pathway for equilibration by exchange scattering. Importantly, the conversion of spin into orbital angular momentum also reduces the net magnetization due to the differing associated Landé g-factors.

Specifically, the z-axis magnetization is $M_z = -g_J \mu_B m_J$, with $g_J$ the Landé g-factor and $m_J$ the magnetic orbital quantum number. The $^6I_J$ multiplet contains several $J, m_J$ states ($J = \frac{7}{2}, \frac{9}{2}, \frac{11}{2}, \ldots \frac{17}{2}$) but these might not all be equally excited. Assuming an equal population of the allowed J states, the net $Gd^{3+}$ magnetization would reduce from $7\mu_B$ ($^8S_{\frac{7}{2}}$) to $6.5\mu_B$ solely due to the optical transition. This ultrafast magnetization reduction is expected to take place *during* the 87 fs-pump excitation, as is consistently seen in the on-resonance measurement ($\tau_d$ (Gd)=38 fs). The population of the $^6I_J$ manifold is furthermore expected to increase, to first order linearly with the pump laser intensity, which we have verified (see Methods). A further argument is found in the very slow remagnetization behavior of Gd. Whereas Fe demagnetizes and remagnetization starts after a few hundred femtoseconds, Gd does not notably remagnetize on a ps time scale (Fig. 3a). Angular momentum and energy would need to be exchanged between the localized 4f multiplet and its surrounding. The excited $^6I_J$ state could non-radiatively relax to another multiplet configuration, $^6P_J$ , at 310 nm (4.02 eV) [24], but this is a forbidden transition and requires interaction with phonons, magnons or itinerant electrons to lose energy. The $^6P_J$ configuration could radiatively relax back to the ground state $^8S_{\frac{7}{2}}$ but this transition is also forbidden. Consequently, the excited $^6I_J$ multiplet is expected to decay slowly, consistent with our measurement.

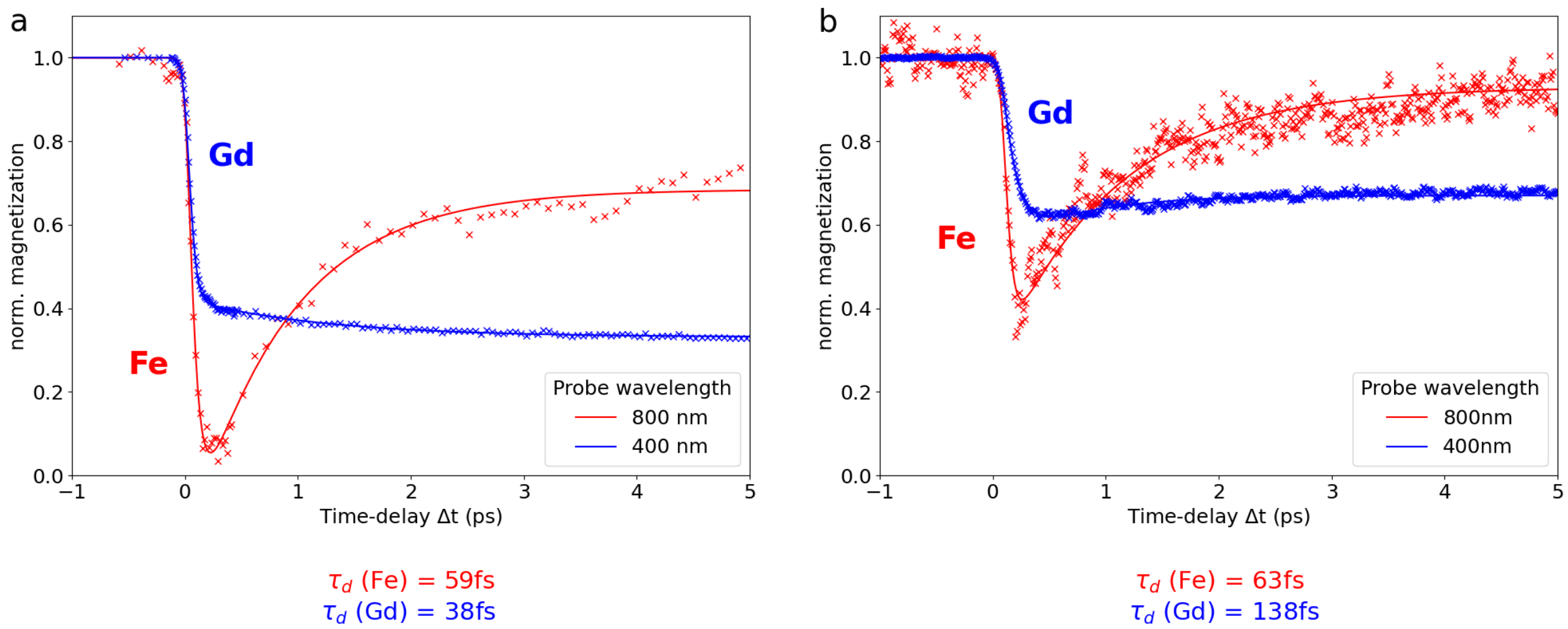

FIG. 3: Normalized total time-resolved magnetization of the two sublattices of GdIG for (**a**) 4f-4f resonant excitation and (**b**) off resonant excitation. The crosses represent the measured data points, while the solid lines correspond to double exponential fits.

While these three features support a direct optical transition to the $^6I_J$ multiplet as mechanism of the exceptionally fast response, the measured $Gd^{3+}$ demagnetization is larger than the estimated magnetization reduction. There can be two reasons for this. There might be an unequal population excited of $J, m_J$ states, which is difficult to assess, since we don't have specific information on their occupations. Second, the magneto-optical constants of the $^8S_{\frac{7}{2}}$ and $^6I_J$ atomic configurations could be distinct, giving the $^6I_J$ configuration a larger apparent demagnetization.

To validate our interpretation of the demagnetization mechanism in GdIG, we conducted a control experiment using a femtosecond pump pulse tuned off-resonance at 340 nm (3.65 eV), as opposed to the resonant excitation at 267 nm (4.65 eV). This excitation wavelength was chosen to lie between the lowest trivalent $Gd^{3+}$ 4f–4f transition $^8S_{\frac{7}{2}} \rightarrow {}^6P_J$ at 310 nm and the (2.86 eV) bandgap.

This off-resonant excitation accesses a different subset of electronic transitions: while it still induces a fraction of 4f–4f intra-atomic excitations in $Gd^{3+}$, it predominantly excites transitions involving the Gd 5d and 6s states, along with transitions in the Fe sublattice. This spectral detuning reduces the resonant 4f multiplet excitation pathway, and it opens other demagnetization and remagnetization channels. Excitations involving Gd 5d and 6s states must exchange angular momentum with the localized 4f states to demagnetize these [16]. Consistent with this scenario, we observe in Fig. 3b that the demagnetization time of the Gd sublattice increases substantially and is with 138 fs a factor of 3.6 slower than the direct inter-multiplet pathway (38 fs). Conversely, the demagnetization time of the Fe sublattice of 63 fs exhibits only a marginal increase, reflecting the persistence of magnon and phonon spin-flip scattering and spin–orbit coupling processes that remain active and, because of the larger Fe bandwidth, do not depend much on the photon energy (267 vs. 340 nm). This result strongly supports our hypothesis that the ultrafast demagnetization of Gd in GdIG is dominantly driven by the direct excitation of 4f multiplet states.

## CONCLUSION

Promoting Gd into excited multiplet states surpasses the common spin-conserving optical selection rules and provides a route to modify the Gd 4f moment during the pump pulse. Such process is limited by the optical transition time and hence appreciably faster than known demagnetization mechanisms that depend on scattering processes occurring after optical excitation, as magnon, phonon, and 5d-4f exchange scattering that are activated by spin-orbit coupling. The intra-multiplet 4f-4f excitation mechanism speeds up the Gd demagnetization by two orders of magnitude, compared to experiments on Gd that employed 800 nm pump pulses and recorded decay times of 14 to 40 ps [7, 16]. The 4f-4f excitation mechanism is one order of magnitude faster than that observed in GdFeCo alloys, wherein the fast demagnetization on the Fe sublattice pulls the Gd moment to revert its direction within one picosecond [14, 32].

The pronounced asymmetry between the ultrafast demagnetization and the substantially slower remagnetization of $Gd^{3+}$ following resonant excitation of the 4f states can be understood in terms of a transient enhancement of spin–orbit coupling. In the electronic ground state of $Gd^{3+}$, the orbital angular momentum is quenched ($L = 0$). Resonant excitation of the 4f manifold temporarily lifts this constraint, resulting in a finite orbital moment and thereby enabling strong spin–orbit coupling. During the demagnetization process, angular momentum is transferred from the spin to the orbital degrees of freedom and subsequently to the lattice as mechanical angular momentum. In the language of the Landau–Lifshitz–Gilbert equation, this corresponds to a transient increase of the effective Gilbert damping parameter.

The reverse process is fundamentally different. Once the excited 4f states have relaxed, the system returns to its ground-state configuration with $L = 0$, closing the spin–orbit-mediated channel for angular momentum transfer. Although the lattice retains the mechanical angular momentum acquired during demagnetization, there is no efficient pathway for transferring it back to the spin system. Consequently, the remagnetization proceeds on a much slower timescale.

The essential mechanism is therefore the transient opening of an additional angular-momentum transfer channel during optical excitation. This channel exists only while the resonantly excited 4f states sustain an enhanced spin–orbit interaction and disappears once the excitation has decayed. As a result, the demagnetization is strongly accelerated, whereas the remagnetization reverts to the intrinsically slower dynamics characteristic of Gd under non-resonant excitation, where the orbital moment remains quenched throughout.

The distinct sublattice dynamics observed under resonant and off-resonant excitation provide clear evidence that 4f multiplet excitations actively govern the demagnetization process. More broadly, our results establish localized 4f excitations as a key lever for controlling 4f magnetization in rare-earth ferrimagnets, opening a route toward tailoring ultrafast magnetic relaxation through selective optical driving.

These findings reveal new possibilities for controlling magnetism in states far from equilibrium. By tailoring the coupling between light and matter at the level of individual electronic subsystems, it becomes possible to engineer ultrafast responses in complex magnetic materials. More broadly, this approach highlights the importance of excitation pathways as an independent control parameter in non-equilibrium condensed-matter physics.

## REFERENCES

* pherrgen@rptu.de

## ACKNOWLEDGEMENTS

Financial support provided by the Deutsche Forschungsgemeinschaft (DFG, German Research Foundation) under Grants No. 318592081 and No. 328545488 (TRR 227, project MF) are gratefully acknowledged. This work was further supported by the Swedish National Infrastructure for Computing (SNIC) funded by the Swedish Research Council through Grant No. 2018-05973. This work has additionally been supported by the Swedish Research Council (VR) and the Knut and Alice Wallenberg Foundation (Grants No. 2022.0079 and No. 2023.0336).

## AUTHOR CONTRIBUTIONS

P.H., U.N., M.Al. and M.Ae conceived the idea and proposed and designed the model and experimental design. B.S., M.Ae. supervised and P.H., T.M. carried out the magneto-optical measurements. C.H. and M.Al. prepared and characterized the GdIG sample. P.H., U.N., P.M.O. and M.Ae. proposed the theoretical explanation. P.H., P.M.O. and M.Ae. co-wrote the manuscript. All authors contributed to the writing of the paper.

## COMPETING INTERESTS

The authors declare no competing interests.

## METHODS

### Sample and Sample preparation

In general, the rare-earth iron garnets ($RE_3Fe_5O_{12}$) consist of three individuals magnetic sublattices, as shown in Fig. 4. The iron atoms in the tetrahedral sites (d) exhibit strong antiferromagnetic exchange coupling with the iron atoms in the octahedral sites (a). Due to their strong exchange parameter $J_{ad} = 22.5\,\mathrm{cm}^{-1}$ [33], they can be treated as a single sublattice. With a ratio of three d-site to two a-site atoms, the iron sublattice has a resulting magnetization in the direction of the d-site iron atoms. The rare-earth element, gadolinium in this case, sits in the dodecahedral sites (c) and also couples antiferromagnetically to the resulting iron sublattice. With an exchange parameter of $J_{cd} = 2.86\,\mathrm{cm}^{-1}$ [34] this exchange coupling is much weaker than the iron-iron coupling.

In our measurements, we investigated a 38 nm-thick gadolinium iron garnet ($Gd_3Fe_5O_{12}$, short: GdIG) film on a (111)-oriented $Gd_3Sc_2Ga_3O_{12}$ (GSGG) substrate. The thin film was grown by pulsed laser deposition using a KrF excimer laser with a fluence of 3–4 J/cm$^2$ and a repetition rate of 5 Hz at room temperature in an oxygen atmosphere at a preasure of $10^{-4}$ mbar. Ex-situ annealing at 600 °C for 30 minutes leads to full crystallisation of the film [35]. The GSGG substrate was chosen to provide a lattice mismatch, which leads to an out-of-plane magnetic easy axis of the GdIG film at room temperature. At ∼130 K, there is a phase transition whereby the magnetic easy axis becomes in-plane below that temperature. The magnetic compensation temperature of the sample is ∼240 K.

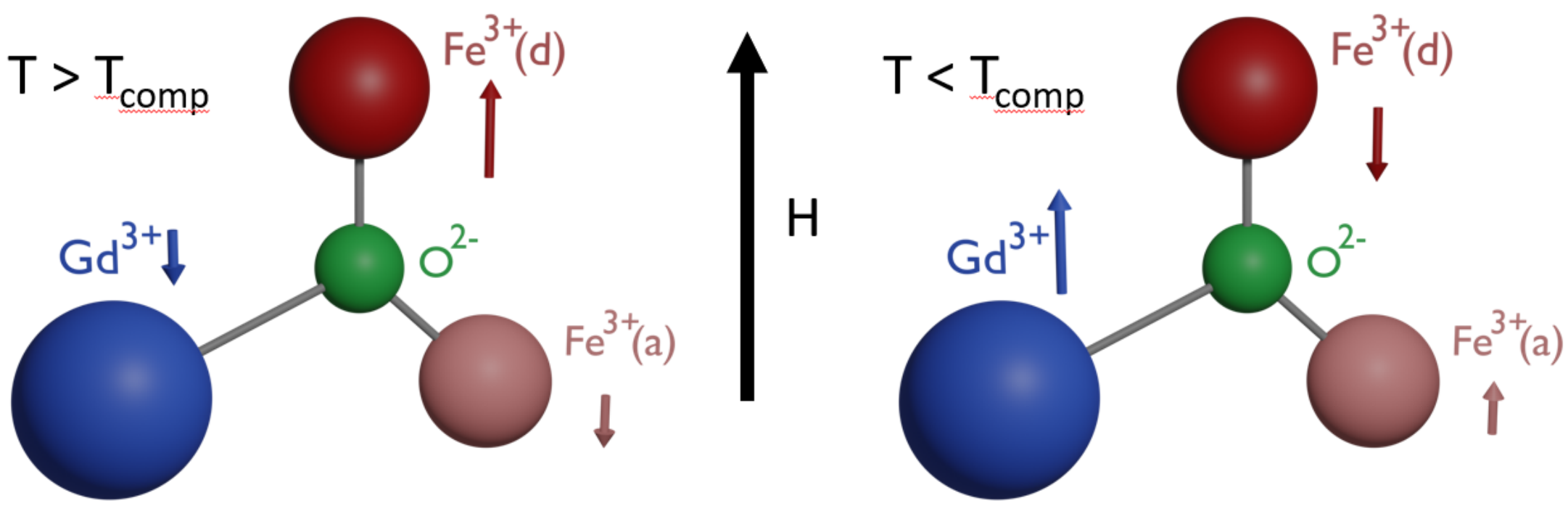

FIG. 4: Magnetic sublattices of GdIG above (left) and below (right) the compensation temperature. For both pictures the external magnetic field is pointing upwards.

### Complex TR-MOKE Method

A pump-probe setup was used to measure these ultrafast magnetization dynamics. Due to the out-of-plane magnetization of the sample, we used polar MOKE geometry. The ultrashort laser pulses (30 fs) are generated by a Ti:Sa amplifier system with a repetition rate of 1 kHz and a fundamental wavelength of 800 nm ($\hat{=}$ 1.55 eV). To overcome the ∼2.86 eV bandgap [21] of the sample we used the third harmonic of our fundamental laser (267 nm $\hat{=}$ 4.65 eV) as the pump wavelength. For probing we used either the fundamental or its second harmonic (400 nm $\hat{=}$ 3.1 eV). By including a quarter-wave plate directly in front of the detector we are able to switch between measuring the Kerr rotation $\theta$ and the Kerr ellipticity $\varepsilon$.

Our first studies on GdIG have revealed distinct temporal behaviors of the Kerr rotation and Kerr ellipticity as functions of time delay for the two magnetic sublattices, as illustrated in Fig. 5. This differential response, while not uncommon in ferrimagnetic systems, highlights a key experimental challenge: the need to account for both the rotation and ellipticity components of the magneto-optical signal in order to fully resolve the dynamics of each sublattice. Unlike simple ferromagnetic materials, where the magnetization can often be inferred from a single parameter, the coexistence of antiferromagnetically coupled $Gd^{3+}$ and $Fe^{3+}$ sublattices in GdIG necessitates a more sophisticated analysis. Specifically, the complex magneto-optical Kerr effect (CMOKE) must be employed, which requires the simultaneous measurement of both Kerr rotation ($\theta$) and Kerr ellipticity ($\varepsilon$) to accurately reconstruct the total magnetization dynamics of the system [36].

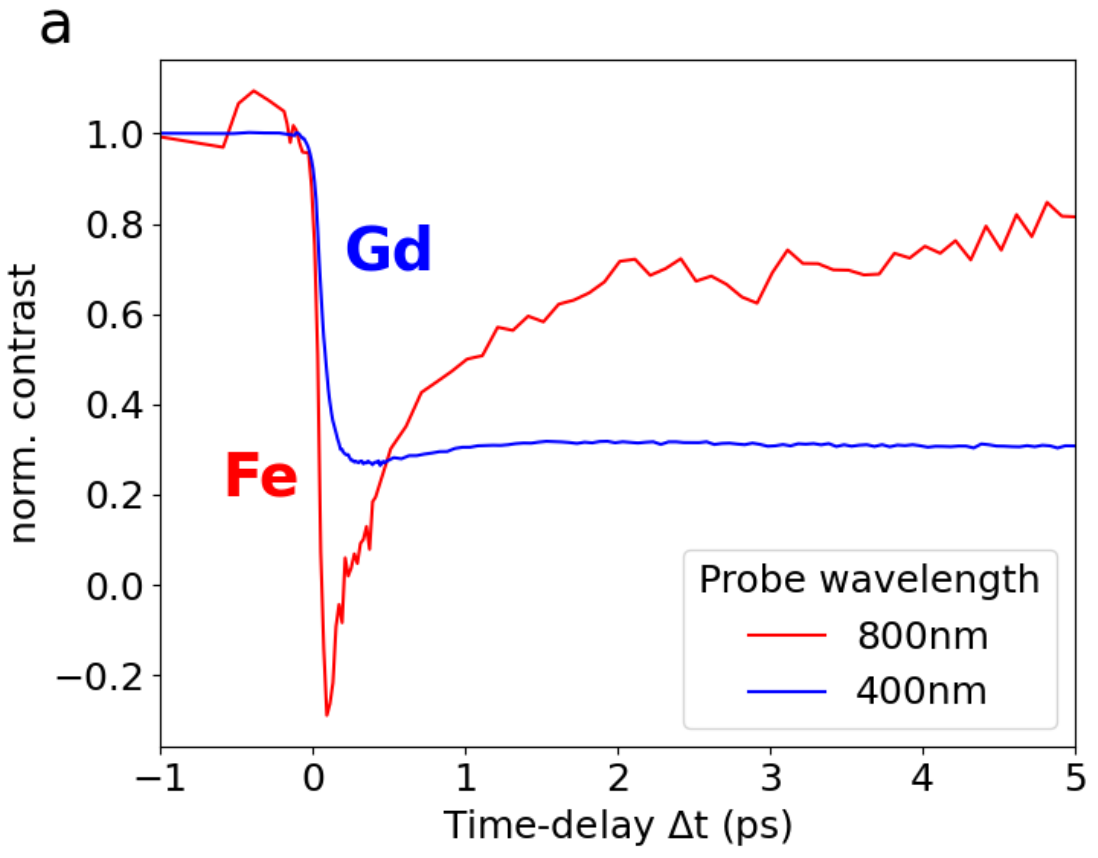

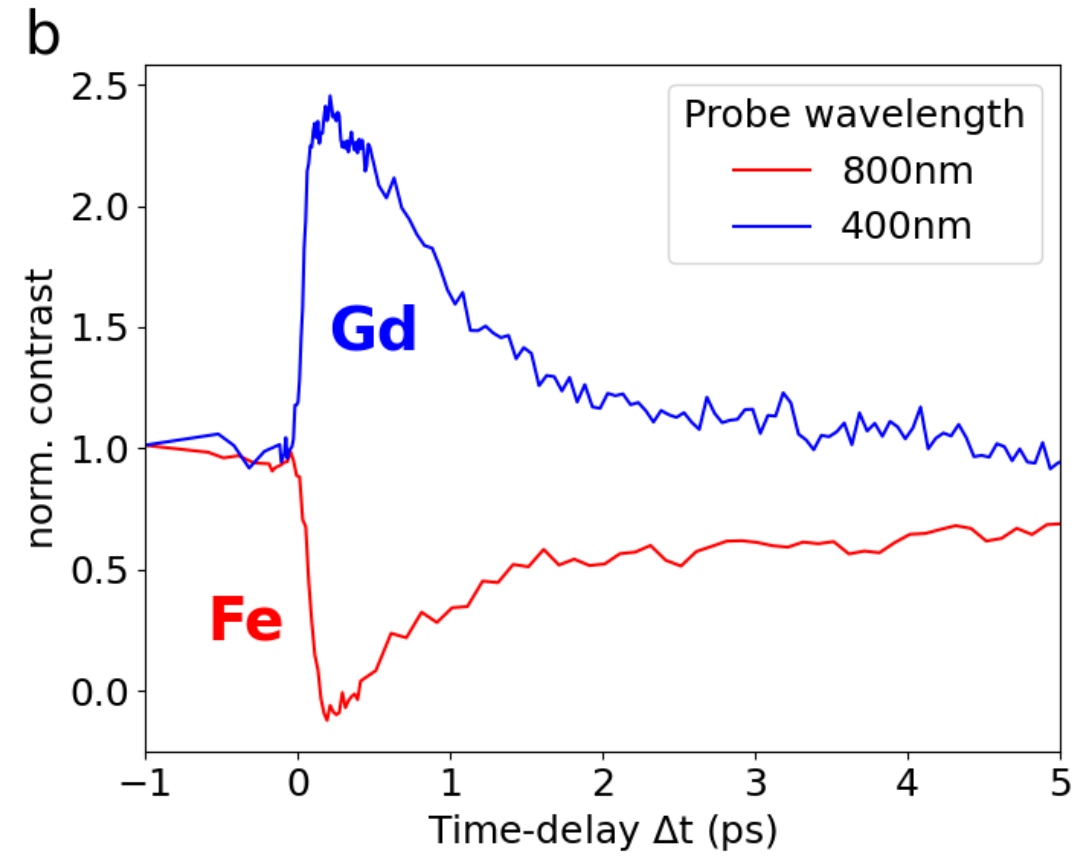

FIG. 5: Kerr-rotation (a) and Kerr-ellipticity (b) of GdIG after a similar optical excitation. Both measurements were performed with a pump wavelength of 267 nm and a pump power of 8 mW.

The combined information from these two quantities can be represented by the complex Kerr vector $\Theta = \theta + i\varepsilon = |\Theta| \cdot e^{i\alpha}$, where $i$ is the imaginary unit and $\alpha$ denotes the phase angle of the vector in the complex plane (see Fig. 6). This formulation allows for a geometric description of the magneto-optical response, with the magnitude $|\Theta|$ representing the total signal strength and the angle $\alpha$ encoding the relative contributions of rotation and ellipticity. For a multilayer or multicomponent magnetic system such as GdIG, the total complex Kerr vector can be approximated as the vector sum of contributions from each magnetic sublattice, each with its own characteristic amplitude and phase. While the intrinsic magnetic response of each layer - reflected in the angle of its individual Kerr vector - remains time-independent, the overall system response evolves dynamically due to differences in the demagnetization rates of the Gd and Fe sublattices. These differences can lead to a time-dependent reorientation of the total complex vector, even when the individual sublattice responses are stable.

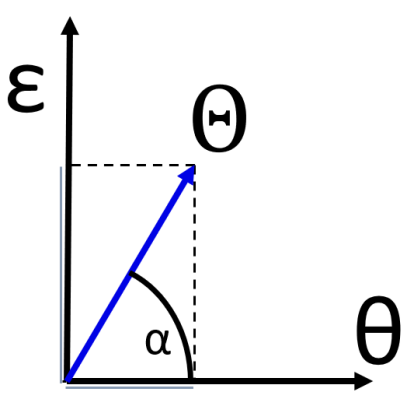

FIG. 6: Schematic illustration of the complex Kerr vector Θ.

Despite this dynamic evolution, the total magnetization of the system can be consistently tracked by computing the magnitude of the total complex Kerr vector, given by $|\Theta| = \sqrt{(\theta^2 + \varepsilon^2)}$. This scalar quantity provides a robust measure of the net magnetic signal, independent of the phase differences between the sublattices. This approach, therefore, enables a quantitative and physically meaningful extraction of the magnetization dynamics of both sublattices from the measured optical response, even in the presence of complex interference effects. The use of CMOKE thus represents a critical methodological advancement for studying ultrafast dynamics in ferrimagnetic garnets, ensuring that the full complexity of the spin system is captured in a way that is both rigorous and experimentally accessible.

## REFERENCES

* pherrgen@rptu.de